\documentclass{iau} 
\usepackage{graphicx}
\usepackage{natbib}

\newbox\grsign \setbox\grsign=\hbox{$>$} \newdimen\grdimen \grdimen=\ht\grsign
\newbox\simlessbox \newbox\simgreatbox
\setbox\simgreatbox=\hbox{\raise.5ex\hbox{$>$}\llap
     {\lower.5ex\hbox{$\sim$}}}\ht1=\grdimen\dp1=0pt
\setbox\simlessbox=\hbox{\raise.5ex\hbox{$<$}\llap
     {\lower.5ex\hbox{$\sim$}}}\ht2=\grdimen\dp2=0pt
\def\simgreater{\mathrel{\copy\simgreatbox}}

\newbox\simppropto
\setbox\simppropto=\hbox{\raise.5ex\hbox{$\sim$}\llap
     {\lower.5ex\hbox{$\propto$}}}\ht2=\grdimen\dp2=0pt

\title[IAUS 351 - Is the Galaxy a typical galaxy?] 
{The building blocks of the Milky Way halo using APOGEE and Gaia \\
	or \\
Is the Galaxy a typical galaxy?}

\author[Schiavon et al.]   
{Ricardo P. Schiavon$^1$, J.Ted Mackereth$^{1,2}$, \\
Joel Pfeffer$^1$, Rob A. Crain$^1$ \& Jo Bovy$^3$}

\affiliation{$^1$Astrophysics Research Institute, Liverpool John Moores
University, Liverpool, L3 5RF, UK \\ R.P.Schiavon@lmu.ac.uk \\[\affilskip]
$^2$Dept. of Astrophysics and Astronomy, Birmingham University, 
Birmingham, UK\\[\affilskip] $^3$Dept. of Astronomy and Astrophysics,
University of Toronto, Toronto, Canada}

\pubyear{2019}
\volume{351}  
\jname{Star Clusters: From the Milky Way to the Early Universe}
\editors{A. Bragaglia, M.B. Davies, A. Sills \& E. Vesperini, eds.}
\begin{document}

\maketitle

\begin{abstract}
We summarise recent results from analysis of APOGEE/Gaia data for stellar
populations in the Galactic halo, disk, and bulge, leading to constraints
on the contribution of dwarf galaxies and globular clusters to the stellar
content of the Milky Way halo.  Intepretation of the extant data in light
of cosmological numerical simulations suggests that the Milky Way has been
subject to an unusually intense accretion history at $z\simgreater1.5$.
\end{abstract}

\firstsection 
\section{Introduction}

The constitution of the Milky Way (MW) halo is one of the foundational
unknowns of the field of Galactic Archaeology \citep[GA,][]{els62,sz78}.
Fundamentally one wants to know how much of the halo was formed
{\it in situ}, how much was accreted, and the detailed properties
of the {\it in situ} and accreted populations, namely masses,
chemical compositions, stellar ages, IMF.  Answering these questions
will require determining the accretion history of the MW.  As it
turns out, accretion histories are a fundamental prediction of the
$\Lambda$-CDM paradigm of structure formation, which opens up the
fascinating perspective that GA can possibly pose tests to galaxy
formation theory.

However, in order for observations of the MW to be deeply consequential
in our pursuit of understanding structure formation, the defining
properties of the MW must be representative of those of other
galaxies.  We are approaching a time where the question of whether
the MW is a typical galaxy can be addressed, on account of a
confluence of the following factors: the revolutionary growth of
data on the MW stellar populations afforded by Gaia and massive
spectroscopic surveys, large surveys of the global properties of
galaxies in a range of redshifts, and the emergence of cosmological
simulations that yield broadly realistic galaxy populations.  The
question is nonetheless not a simple one to formulate (e.g., typical in
which respect?) and it surely will not be an easy one to answer.
Nevertheless, it is crucial that we ask it, as it has far-reaching
implications.  From a broad perspective, the question ultimately
has a bearing on the cosmic history leading up to our presence in
the universe.  
In a more restricted context 
the power of MW data to constrain galaxy formation theory
depends on whether it is representative of its similarly-massive peers,
or an outlier.

In this paper we summarise three lines of evidence, from studies
of the halo, disk, and bulge, suggesting that the Galaxy may not
after all be a typical galaxy.  The work by our group summarised
in this paper was all based on analysis of data from SDSS-IV/APOGEE
\citep{Majewski2017}, Gaia DR2 \citep{GaiaDR2}, and the EAGLE
cosmological numerical simulations \citep{Schaye2015,Crain2015}.

\section{The Galactic Halo}
\label{halo}

An exciting recent result in GA was the discovery by different
groups that the local stellar halo is dominated by the remnants of
the accretion of a single satellite galaxy
\citep{Haywood2018,Helmi2018,Belokurov2018,Myeong2018,Mackereth2019}.  The
Gaia-Enceladus/Sausage (GE/S) system is thought to be a dwarf galaxy
which merged with the MW $\sim$~10~Gyr ago.  It was identified
as a relatively metal-poor, low-$\alpha$ stellar population dominated
by highly eccentric and slightly retrograde orbits.  It is also
characterised by low [Ni/Fe], which is a typical abundance trait
of nearby satellite galaxies \citep[e.g.,][]{Shetrone2003}.  Based on the
width of its metallicity distribution, \cite{Helmi2018} estimated the mass
of GE/S to be $\sim~10^9$M$_\odot$.

Further constraints on the mass of GE/S and the time of accretion
were obtained from an examination of predictions by the EAGLE
simulations.  \cite{Mackereth2019} contrasted the chemical compositions
and orbital properties of GE/S with those of their analogues in the
L025N752-Recal ($25^3$ Mpc$^3$) EAGLE simulation, constraining 
$M^\star$ to be around $\sim~10^{8.5}-10^9$M$_\odot$
on the basis of the mean [Fe/H] and [Mg/Fe] of GE/S stars.  Moreover,
they showed that the highly eccentric nature of GE/S stars imply
that the merger must have occurred no earlier than $z\sim1.5$
\cite[][Fig.~9]{Mackereth2019}.  That is because the maximum impact
parameter for a successful accretion depends on the gravitational
potential of the central halo.  At higher $z$, when central galaxies
were smaller and less massive, mergers with small enough impact
parameters that result in highly eccentric orbits were too rare.

Interestingly, an analysis of the accretion histories
of galaxies in the $25^3$ Mpc$^3$ volume
showed that only $\sim10\%$ of MW-like galaxies underwent an accretion
event similar (in terms of mass and timing) to that of GE/S.  By
considering other ongoing and suspected accretion events, such as
Sgr dSph, LMC, SMC, and the Kraken \citep{Kruijssen2018}, one would
conclude that the accretion profile of the MW may be even more
uncommon.

\section{Disk $\alpha$-bimodality and the accretion history of the MW}
\label{disk}

One of the most puzzling features of the MW disk is its so-called
$\alpha$-bimodality, which is defined as the occurrence, in the
same location in the Galaxy, of two stellar populations with
substantially different [$\alpha$/Fe] and a sizeable overlap in
[Fe/H] \citep[see, e.g., Fig.~22 of][]{Bensby2014}.  The phenomenon
extends across most of the MW disk \citep[][Fig.~4]{Hayden2015}.
In addition, the high-$\alpha$ component has a longer scale height
and shorter scale length than its low-$\alpha$ counterpart
\citep{Bovy2012,Mackereth2018}.  These observations are very hard
to explain on the basis of standard Galactic chemical evolution
models \citep[e.g.,][]{Andrews2017}.  In order to connect two stellar
populations with same [Fe/H] and different [$\alpha$/Fe] by a star
formation and chemical enrichment path following the laws of chemical
evolution, such models often need to resort to {\it ad hoc} assumptions
on, e.g., gas inflows and star formation efficiency \citep[but
see][]{Chiappini2009,Schonrich2009,Clarke2019}.

\cite{Mackereth2018} analysed the EAGLE simulations
in order to gain insights into the physical motivation behind the
$\alpha$-bimodality.  They identified 133 MW-like galaxies in the
Ref-L100N1504 ($100^3$ Mpc$^3$) EAGLE simulation on the basis of
their masses and (kinematically defined) morphologies.  After careful
scrutiny of the distribution of stellar populations in the simulated
MW-like galaxies on the [Fe/H] vs.\ [Mg/Fe] plane, only six galaxies
were found to display some type of $\alpha$-bimodality in their
disk populations.  Tracing back in time the evolution of the gas
particles leading up to the formation of the low- and high-$\alpha$
stellar populations, it was found that they evolved in complete
chemical detachment.  In other words, they never exchanged gas and
thus there is no need to devise a star formation and chemical
enrichment path connecting populations with different [$\alpha$/Fe].
Instead, low-/high-$\alpha$ populations were formed in regions
of long/short gas consumption timescale, which
in turn is regulated by gas pressure.

As mentioned above, only about 5\% of all MW-like galaxies in the
simulated volume display some kind of $\alpha$-bimodality.
Investigating what sets
these few cases apart from the general MW-like galaxies in the simulation,
\cite{Mackereth2018} found that they underwent a different
accretion history.  Simulated galaxies whose
disk populations show $\alpha$-bimodality accreted satellites more
intensely at $z \simgreater 1.5$ than their normal counterparts
\citep[Fig.~9 of][]{Mackereth2018}.  Thus, in agreement with
the stellar halo study described in Section~\ref{halo}, 
evidence from the chemistry of its disk populations suggests
that the MW history has been characterised by atypical accretion
activity.

\section{Globular cluster remnants in the inner Galaxy} 
\label{inner}
While analysing APOGEE DR12 elemental abundances for a sample of
~5,200 bulge stars, \cite{Schiavon2017} discovered
a large population of field stars within $\sim~2$~kpc of the Galactic
centre, characterised by very high N and Al abundances, which are
anti-correlated with the abundance of C.  The chemistry of
these {\it N-rich stars} is typical of the so-called ``second
generation'' stars first identified in globular clusters (GCs)
many years ago \citep[see][]{Bastian2018,Renzini2015}.
The N-rich star metallicity distribution differs to a high
degree of confidence from that of the existing Galactic GC system,
so that N-rich stars cannot be explained by evaporation and tidal
stripping from those GCs.  \cite{Schiavon2017} hypothesised that
N-rich stars result from destruction of an early population
of GCs.  The mass in destroyed GCs within 2 kpc of the Galactic
centre is $\sim~1.5-2 \times 10^8$~M$_\odot$, corresponding to 20-50\%
of the halo stellar mass within that volume, depending on the
destroyed GC and halo mass estimates.  It also amounts to several
times the mass of the existing GC system.  It is possible
that this population was deposited in the inner halo a long time
ago, partly from accreting systems, and partly during {\it in situ}
formation in a turbulent disc, followed by immediate tidal disruption
\citep[e.g.,][]{Kruijssen2015} from interaction with massive molecular
clouds.

The possible presence of a large stellar mass in destroyed GCs in
the heart of the MW is another indication of a very high accretion
rate in early times.  A high accretion rate brings about high gas
pressure, leading to high GC formation and destruction
\citep{Pfeffer2018}, which would in turn cause the MW today to have
a smaller/less massive GC system than average galaxies of the same
halo mass.

But does the MW have a low total GC mass for its halo
mass?   In fact, the total mass in GCs ($M_{GC}$) is perhaps the
best tracer of the halo mass of a galaxy ($M_h$), with a ratio $\eta
= M_{GC}/M_h = 4\times10^{-5}$ that is constant
over several decades in mass, with a scatter of only 0.2 dex
\citep[e.g.,][]{Hudson2014}.  Considering accepted ranges for
$M_{h,MW}$ and $M_{GC,MW}$ \citep{Watkins2019,Hudson2014,Kruijssen2009}
we have $\eta_{MW}=1-4\times10^{-5}$, or somewhere between
average or too low by over 2$\sigma$.


The uncertainties in the relevant quantities are still too large
for a call on whether the early formation/accretion/destruction of GCs
was exceptionally large in the past.  If $\eta_{MW}$ is average,
one would be led to conclude that in general galaxies destroyed GCs just as
vigorously as the early MW, making GCs (or their parent populations) important
contributors to the stellar mass budget at the central regions of all massive
galaxies.  If however $\eta_{MW}$ is significantly low, the conclusion
is again that the MW may have undergone an unusually intense accretion
history in the distant past.

\section{Concluding Remarks} 
\label{conclusions} 

We briefly reviewed recent work taking stock of the contributions
to the stellar mass budget of the MW halo by dwarf galaxies and
globular clusters, based on recent data from cutting edge surveys
of Galactic stellar populations.  Intepretation of these findings
in light of state-of-the-art cosmological numerical simulations
suggests that the accretion history of the MW may have been unusual,
currently placing it in the 5\% category.  It will be interesting
to see how this picture will evolve as a result of the upcoming
ten-fold increase in halo sample size afforded by the next generation of
spectroscopic surveys \citep[e.g., WEAVE, Jin et al.\ 2019, in
prep., Dalton et al. 2016; 4MOST,][]{deJong2019}, complemented by improvements in both
resolution and subgrid physics of numerical simulations.  Maybe
those developments will locate the MW ever farther from the typical
$L^\star$ disk galaxy, or maybe they will shift it back to a position
of normality.  Either way, ten years from today we will have expanded
our knowledge of the make up of the MW halo and the history of our
Galaxy greatly.  These are exciting times.

\end{document}